\documentclass[aps,preprint,nofootinbib]{revtex4}
\usepackage{sublabel}
\usepackage{graphicx}
\usepackage{epsf}
\usepackage{wrapfig}
\usepackage{epsfig}
\def\Vec#1{\mbox{\boldmath $#1$}}
\def\bra#1{\langle#1\,|}
\def\ket#1{|\,#1\rangle}

\def\beq{\begin{equation}}
\def\eeq{\end{equation}}

\def\beqy{\begin{eqnarray}}
\def\eeqy{\end{eqnarray}}
\def\ie{i.e.}
\def\eg{\textit{e.g.}}

\def\ma{\hat{A}}
\def\mh{\hat{\mathcal{O}}}

\begin{document}
\vskip 2mm \date{\today}\vskip 2mm
\title{A new realistic many-body approach for the description
  of high-energy scattering processes off complex nuclei}
\author{M. Alvioli}
\author{C. Ciofi degli Atti}
\address{Department of Physics, University of Perugia and
      Istituto Nazionale di Fisica Nucleare, Sezione di Perugia,
      Via A. Pascoli, I-06123, Perugia, Italy}
\author{H. Morita}
\address{Sapporo Gakuin University, Bunkyo-dai 11, Ebetsu 069-8555,
  Hokkaido, Japan}
\vskip 1cm
%% ======================================================================= %%
\begin{abstract}
A linked cluster expansion for the calculation of ground state observables 
of nuclei with realistic interactions has been developed.
Using the $V8^\prime$
potential the ground state energy, density and momentum distribution of complex nuclei
have been calculated and found to be in good agreement with the results obtained within
the Fermi Hyper Netted Chain, and Variational Monte Carlo approaches. Using the same 
cluster expansion, with wave function and correlations parameters fixed from the
calculation of the ground-state observables, various high energy scattering
processes off complex nuclei have been calculated taking final state interaction
effects into account by means of the Glauber multiple scattering series.
\end{abstract}
\maketitle
%% ======================================================================= %%
\section{Introduction}
The knowledge of the nuclear wave function, in particular its most interesting
and poorly known  part, {\it viz} the correlated one, 
is not only a prerequisite for understanding the details of bound hadronic systems,
but is becoming at present a necessary ingredient for a correct description of
medium and high energy scattering processes off nuclear targets; these in fact
represent a current way of investigating short range effects in nuclei as well
as those QCD effects (\eg color transparency,  hadronization, dense
hadronic matter, etc) which manifest themselves in the nuclear medium.
The problem of using a realistic many-body description is not trivial, for one
has first to solve the many body problem
and then has to find a way to apply it to scattering processes. 
The  difficulty mainly arises because even if  a reliable and manageable 
many-body description of the ground state is developed, the
problem remains of the calculation of the final state.
In the case of \textit{complex nuclei}, much remains to be done in order to achieve
a consistent treatment of Initial State Correlations (ISC) and final state interaction
(FSI), which is feasible in the case of \textit{few-body systems} while to date is
still lacking for nuclei with mass number larger than $A=4$; the structure of the best
trial wave function resulting from very sophisticated
calculations (\eg the variational Monte Carlo ones) is so
complicated, that its use in the calculation of various processes at intermediate
and high energies appears to be not easy task.
We have developed an economic method for the calculation of the ground-state
properties (energies, densities and momentum distributions) of complex nuclei within
a framework which can be easily applied to the treatment of various scattering
processes, keeping  the basic features of ISC as predicted by the structure of
realistic Nucleon-Nucleon (NN) interactions.
%% ======================================================================= %%
\section{The cluster expansion}\label{sectionCLUSTER}

\begin{figure}[!htp]
\centerline{\psfig{file=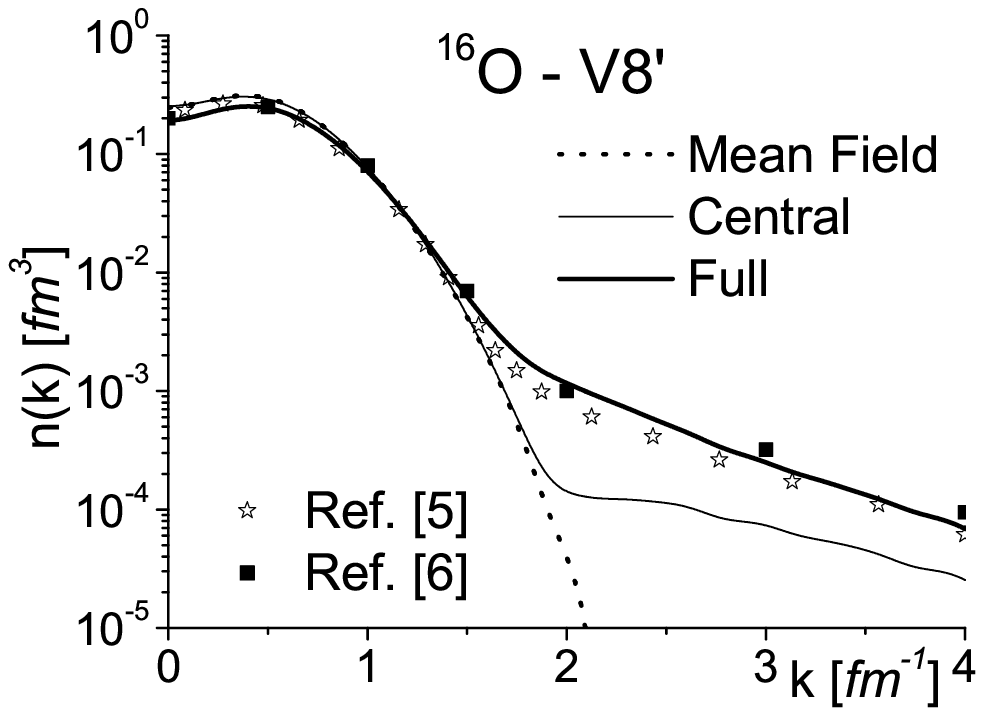,width=6.5cm}\,\psfig{file=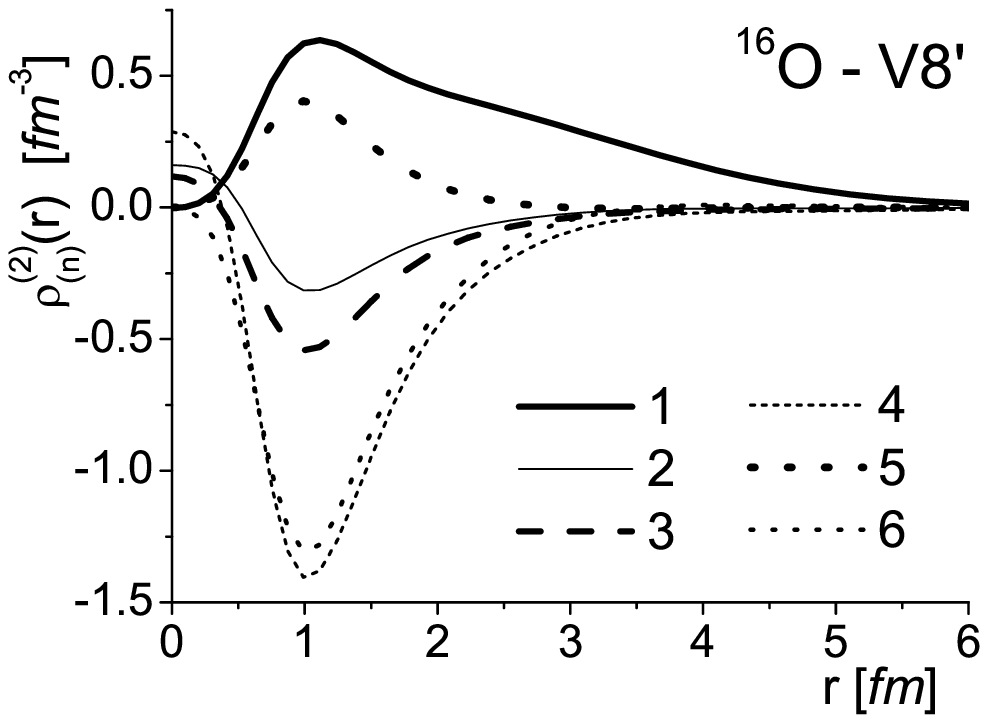,width=6.5cm}}
\vspace*{8pt}
\caption[]{The momentum distribution (Eq. (\ref{defmomdis}); \textit{left}) and the
\textit{radial} two-body density (Eq. (\ref{hamilt}; \textit{right}) of $^{16}O$
calculated within the cluster expansion. 
The $\eta$-expansion (\textit{thick solid}) result for the momentum ditribution is
compared with the VMC~\cite{pie01} (\textit{squares}) and the FHNC~\cite{fab01} (\textit{stars})
ones; the mean field (\textit{dots}) and the Jastrow (\textit{thin solid}) results is also
show, in the left figure. 
The six components of $\rho^{(2)}_{(n)}$~\cite{alv01,alv02}, which couple with the first
six components of the $V8^\prime$ potential are shown in the right figure.}\label{Fig1}
\end{figure}

\begin{figure}[!hbp]
\centerline{\psfig{file=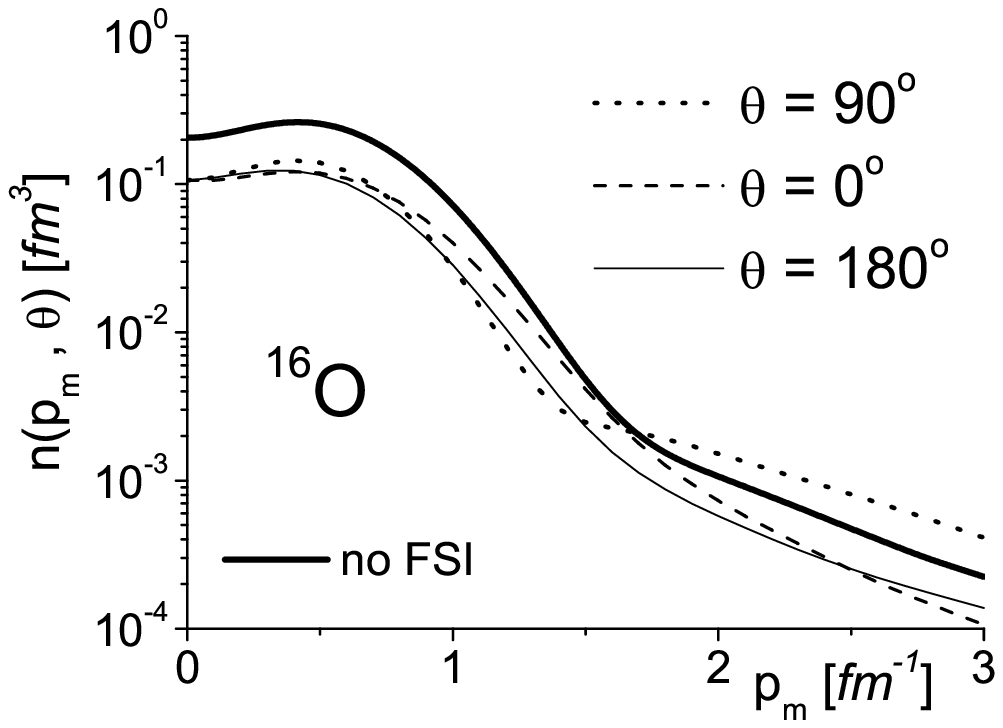,width=6.5cm}\,\psfig{file=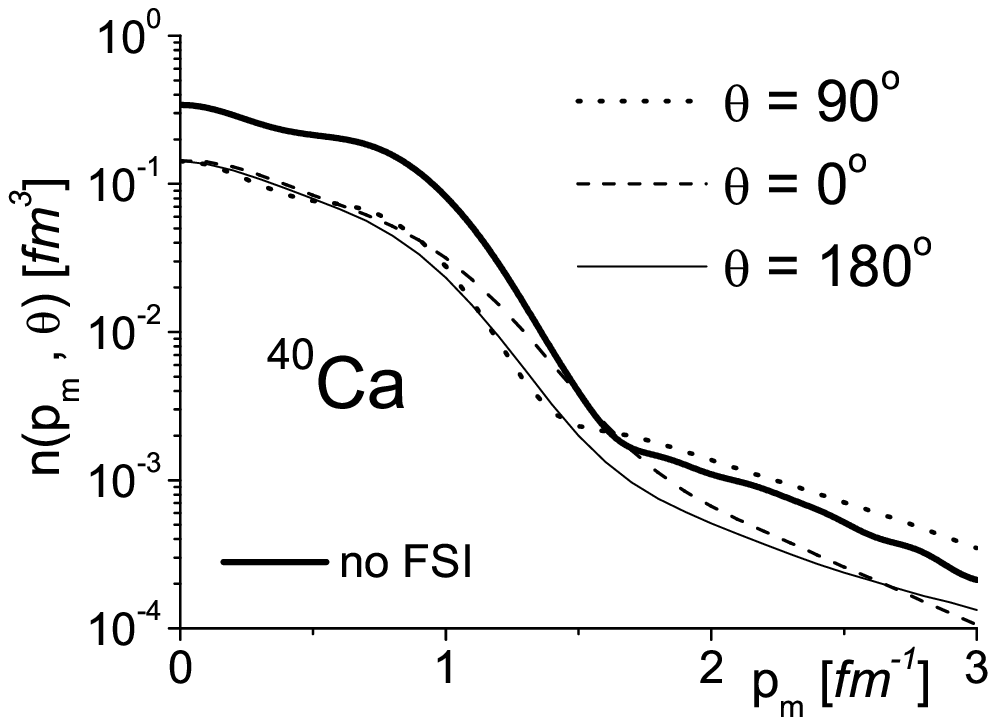,width=6.5cm}}
\vspace*{8pt}
\caption[]{The \textit{distorted} momentum distribution of Eq. (\ref{momdisto}) for
$^{16}O$ (\textit{left}) and $^{40}Ca$ (\textit{right}). \textit{Thick solid} line:
no final state interaction taken into account; \textit{dotted}, \textit{dashed} and
\textit{thin solid} lines: FSI in the Glauber framework for \textit{perpendicular},
\textit{parallel} and \textit{antiparallel} kinematics, respectively; $\theta$ represents
the angle between the missing momentum $\Vec{p}_m$ and the three-momentum transfer,
$\Vec{q}$.}\label{Fig2}
\end{figure}

We write the nuclear Hamiltonian in the usual form, \ie:
\beq
\label{hamilt}
\hat{H}\,=\,\hat{T}\,+\,\hat{V}\,
        =\,-\frac{\hbar^2}{2\,M_N}\,\sum^A_{i=1}\,\nabla^2_i\,
                \,+\,\sum_{i<j}\,\hat{v}_2(\Vec{x}_i,\Vec{x}_j)\;,
\eeq
where the vector $\Vec{x}$ denotes the set of nucleonic degrees of freedom,
and the two-body potential $\sum_{i<j}\,\hat{v}_2(\Vec{x}_i,\Vec{x}_j)
=\sum_{n=1}^{N}\,v^{(n)}(r_{ij})\,\mh^{(n)}_{ij}$, with $r_{ij}=|\Vec{r}_i-\Vec{r}_j|$
denoting the relative distance of nucleons $i$ and $j$, contains the well known
spin- and isospin-dependent operator $\mh^{(n)}_{ij}$. 
The evaluation of the expectation value of the nuclear Hamiltonian (\ref{hamilt})
is object of intensive activity which, in the last few years, has produced
considerable results;  our goal is to present an economical, but effective
method for the calculation of the expectation value of any quantum
mechanical operator $\ma$
\beq
\label{omedio1}
\langle\,\ma\,\rangle\,=\,\frac{\bra{\psi_o}\,\ma\,
        \ket{\psi_o}}{\bra{\psi_o}\psi_o\rangle}\,
        =\,\frac{\bra{\phi_o}\,\hat{F}^\dagger\,\ma\,\hat{F}\,
        \ket{\phi_o}}{\bra{\phi_o}\,\hat{F}^\dagger\,\hat{F}\,\phi_o\rangle}\,;
\eeq
where $\phi_o$ is a Shell-Model (SM),  mean-field wave function, and $\hat{F}$ is 
a symmetrized \textit{correlation operator}, which generates correlations
into the mean field wave function. According to the two-body interaction in Eq.
(\ref{hamilt}), the correlated ground state wave function $\phi_o$ is cast in the
following form:
\beq
\label{newdue}
\hat{F}(\Vec{x}_1,\Vec{x}_2\,...\,\Vec{x}_A)\,\phi_o(\Vec{x}_1,\Vec{x}_2\,...\,\Vec{x}_A)\,
=\,\hat{S}\,\prod^A_{i<j}\,\hat{f}(r_{ij})\,\phi_o(\Vec{x}_1,\Vec{x}_2\,...\,\Vec{x}_A)
\eeq
%%%---------------------
\begin{table}[!htp]
\caption{Benchmark calculation for the ground state potential, kinetic, total energy
and energy per particle for $^{16}O$, at the first order of $\eta$-expansion
(Eq. (\ref{seventeen})), compared with the FHNC results for the corresponding
quantities calculated with the same wave function.}
{\begin{tabular*}{\textwidth}{@{}c||cccc@{}} \toprule
  &{\bf V}&{\bf T}&$\mathbf{E_o}$&$\mathbf{E_o/A}$\\\hline
$\eta-$\textit{exp}~\cite{alv01} & -390.37& 323.50& -66.87 &-4.18\\\hline
$FHNC$~\cite{fab01} & -390.30& 325.18& -65.12 &-4.07\\\hline
\end{tabular*}\label{Tab1}}
\end{table}
%%%---------------------
with
\beq
\hat{f}(r_{ij})=\sum_{n=1}^{N}\,\hat{f}^{(n)}(r_{ij})\;\;\hspace{2cm}
        \hat{f}^{(n)}(r_{ij})=f^{(n)}(r_{ij})\,\hat{O}^{(n)}_{ij}\,.
\label{corrop1}
\eeq
In the following, we are going to describe the cluster expansion ($\eta-expansion$)
technique we used to evaluate Eq. (\ref{omedio1}); the solution can be found by
applying the variational principle, with the variational parameters contained both
in the correlation functions and in the mean field single particle wave functions.
The expectation value $\langle \hat{A} \rangle$ defined in (\ref{omedio1}) can
be cluster-expanded; at first order of the $\eta-$expansion~\cite{alv01,alv02,alv03},
it reads as follows
\beq
\label{seventeen}
\langle\hat A\rangle_1=\bra{\phi_o}\sum_{i<j}\left(\hat{f}_{ij}\hat A
        \hat{f}_{ij}-\hat{A}\right)\ket{\phi_o}
      -\langle \hat A\rangle_o\bra{\phi_o}\sum_{i<j}
      \left(\hat{f}_{ij}
        \hat{f}_{ij}-1\right)\ket{\phi_o}\,,
\eeq
where $\langle \hat A\rangle_o$ is given by $\bra{\phi_o} \hat{A}\ket{\phi_o}$.
The $2nd$ order term can straightforwardly be obtained by the same technique
used to derive Eq. (\ref{seventeen}).
Given the two-body interaction as in Eq. (\ref{hamilt}), the expectation
value  of the Hamiltonian can be written as:
\beq
\label{hmatrix}
E_o\,=\,\int\,d\Vec{k}\,k^2\,n(\Vec{k})\,+\,\sum_n\;\int\;d\Vec{r}_1 d
        \Vec{r}_2\;v^{(n)}(r_{12})\rho^{(2)}_{(n)}(\Vec{r}_1,\Vec{r}_2)\,,
\eeq
where $\rho^{(2)}_{(n)}(\Vec{r}_1,\Vec{r}_2)$ is the (spin and isospin dependent-)
Two Body Density (TBD) matrix~\cite{alv01,alv02}, and $n(k)$ is the nucleon
momentum distribution, defined in terms of the non-diagonal One Body Density Matrix
(OBDM),  $\rho^{(1)}(\Vec{r}_1,\Vec{r}_1^\prime)$, as
\beq
\label{defmomdis}
n(\Vec{k})\,=\,\frac{1}{(2\pi)^3}\,\int d\Vec{r}_1 d\Vec{r}^\prime_1
        \,e^{i\,\Vec{k}\cdot(\Vec{r}_1-\Vec{r}^\prime_1)}\;\rho(\Vec{r}_1,
        \Vec{r}^\prime_1)\,.
\eeq
The knowledge of the OBDM allows one to calculate other relevant quantities like \eg the 
density distribution $\rho(\Vec{r})\,=\,\rho^{(1)}(\Vec{r}_1=\Vec{r}_1^\prime\equiv\Vec{r})$.
The results of calculation of the ground state energy, the charge density and
the two-body density and momentum distribution using the realistic $V8^\prime$
interaction~\cite{pud01} is discussed in detail in Refs. ~\cite{alv01,alv02,alv03};
in Table \ref{Tab1} some results for a benchmark calculation of the quantity in Eq. 
(\ref{hmatrix}) are shown and it can be seen that the agreement with the reference
FHNC calculation is very good. Results for the momentum distribution and \textit{radial}
TBD, defined as  $\rho^{(2)}_{(n)}(r)=\int d\Vec{R}\,\rho^{(2)}_{(n)}(\Vec{r}_1,\Vec{r}_2)$,
where $\Vec{r}$ and $\Vec{R}$ are the relative and center of mass vectors of particles $1$
and $2$~\cite{alv01,alv02}, are shown in Fig. \ref{Fig1}; 
we obtained the best wave functions for $^{16}O$ and $^{40}Ca$, used to calculate the
quantities shown in Figs. \ref{Fig1} and \ref{Fig2}
and to be used in various applications, from the results achieved variationally
within the FHNC~\cite{fab01} approach, \ie, we used the correlation functions provided 
by the FHNC method and then optimized the single particle wave fucntions in order
to reproduce the experimentally observed charge densities and radii for these two 
nuclei; we found that $^{16}O$ can be described both with with harmonic
oscillator or Saxon-Woods mean field functions while, in the case of $^{40}Ca$, Saxon-Woods
functions have to be used to produce satisfactory densities and momentum distributions,
though the agreement with the FHNC results for the ground state energy is not
as satisfactory as in the $^{16}O$ case~\cite{alv01,alv02}.
%%%---------------------
%%%---------------------
The method we have developed  appears to be a very effective, transparent and parameter-free
one. An extensive discussion of the applications of the method of the cluster expansion 
to scattering problems is out of the scope of the present report and will be presented
elsewhere.
%%%---------------------
%%%---------------------
In this report, we show what the effect of the Glauber final state
interaction~\cite{cio01,mor01} on the distorted nucleon momentum distribution which
appear in the semi-inclusive $A(e,e^\prime p)X$ reaction; \ie
\begin{equation}
\label{momdisto}
n_D({\bf p}_m)\,=\,\frac{1}{(2 \pi)^3}\,\int e^{i {\bf p}_m\,\cdot\,({\bf r}_1 -{\bf r}_1')}
      \rho_D ({\bf r}_1,{\bf r}_1')\, d{\bf r}_1 d{\bf r}_1'\,,
\end{equation}
where $\Vec{p}_m=\Vec{q}-\Vec{p}$ is the missing momentum, defined as the difference
between the three-momentum transfer and the momentum of the detected proton, and
\beq
\label{rodi}
\rho_D ({\bf r}_1,{\bf r}_1')= \frac {\bra{\psi_o}\,S^{\dagger}
\,\hat{\rho}({\bf r}_1,{\bf r}_1')\,S'\,\ket{\psi_o}^\prime}{\bra{\psi_o}
      \,\psi_o\rangle}\
\eeq
is the distorted OBDM. The FSI is introduced by the Glauber approximation, \ie
where $\hat{\rho}_1(\Vec{\tilde{r}}_1,\Vec{\tilde{r}}_1^\prime)$ is the OBDM operator,
$S$ is the Glauber S-matrix, \ie
\begin{equation}
\label{glaudef}
S \rightarrow S_G({\bf r}_1\dots{\bf r}_A)=\prod_{j=2}^AG({\bf r}_1,{\bf r}_j)
      \equiv \prod_{j=2}^A\bigl[1-\theta(z_j-z_1)\Gamma({\bf b}_1-{\bf b}_j)\bigr]
\end{equation}
with asymptotic values of the parameters appearing in the profile $\Gamma$
functions~\cite{cio01}.
The distorted momentum distributions (\ref{momdisto}) are presented in Fig. \ref{Fig2}

\section{summary}

To sum up, we have shown that, using realistic models of the NN interaction,
a proper approach based on cluster expansion techniques  can produce reliable approximations
for those diagonal and non diagonal density matrices which appear in various medium and high
energy scattering processes off nuclei, so that the role of nuclear effects in these
processes can be reliably estimated without using free parameters to be fitted to the data. 
Our approach has been applied to semi-inclusive $A(e,e^\prime p)X$ reaction~\cite{alv02,alv03};
we also calculated the total neutron-nucleus cross section~\cite{alv04,alv05} within the same 
Glauber approach, obtaining, once inelastic shadowing effects are taken into account, a
very good agreement with data over a wide range of mass number A, which confirms
NN correlations can play an inmportant role in high-energy scattering processes.
%% ======================================================================= %%

%% ======================================================================= %%

\begin{thebibliography}{12}
%
\bibitem{alv01} M. Alvioli, C. Ciofi degli Atti, H. Morita,
  \textit{Phys. Rev.} \textbf{C72} 054310 (2005)
%
\bibitem{alv02} M. Alvioli, \textit{PhD Thesis}, \textit{University of Perugia} (2003).
%
\bibitem{alv03} M. Alvioli, C. Ciofi degli Atti, H. Morita,
  \textit{Fizica} \textbf{B13} 585 (2004).
%
\bibitem{pud01} B. S. Pudliner, V. R. Pandharipande, J. Carlson, S. C. Pieper
      and R. B. Wiringa, \textit{Phys. Rev.} \textbf{C56} (1999) 1720.
%
\bibitem{fab01} A. Fabrocini, F. Arias de Saavedra and G. Co', 
  \textit{Phys. Rev.} \textbf{C61}, 044302 (2000) and \textit{Private Communication}.
%
\bibitem{pie01} S.C. Pieper, R.B. Wiringa and V.R. Pandharipande,
      \textit{Phys. Rev.} \textbf{C46}, 1741 (1992).
%
\bibitem{cio01}  C. Ciofi degli Atti, and D. Treleani,  
     \textit{Phys. Rev} \textbf{C60} (1999) 024602.
%
%\bibitem{bra01} M. A. Braun, C. Ciofi degli Atti and D. Treleani,
%      \textit{Phys. Rev.} \textbf{C62} (2001) 034606.
%
\bibitem{mor01} H. Morita, M. A. Braun, C. Ciofi degli Atti and D. Treleani,
      \textit{Nucl. Phys.} \textbf{A699} (2002) 328c.
%
%\bibitem{bro01} S. J. Brodsky and A. H. Mueller,
%      \textit{Phys. Lett.} \textbf{B206} (1988) 685
%
\bibitem{alv04} M. Alvioli, C. Ciofi degli Atti, I. Marchino and H. Morita,
  \textit{in preparation}
%
\bibitem{alv05} M. Alvioli, C. Ciofi degli Atti, I. Marchino and H. Morita;
  invited talk at the \textit{XXIV International Workshop on Nuclear Theory,
    Rila, Bulgaria} (2005); \texttt{nucl-th/0510079}
%
%\bibitem{gri01} V. N. Gribov, \textit{Sov JETP} \textbf{29} (1969) 483;
%  J. Pumplin, M. Ross, \textit{Phys. Rev. Lett.} \textbf{21} (1968) 1778;
%  V. A. Karmanov,L. A. Kondratyuk, \textit{Jetp letters} \textbf{18} (1973) 451.
%
%\bibitem{jen01} B. K. Jennings and G. A. Miller, \textit{Phys. Rev.} \textbf{C49}
%  (1999) 2637.
%
%\bibitem{transp} A. Mueller, in \textit{Prooceedings of the 17th rencontre de Moriond},
%edited by J. Tranh Thanh Van (Edition Frontieres, Gif-sur-Ivette, 1982), p. 13;
%                 S. J. Brodsky, in \textit{Prooceedings of the 13th International
%		 Symposium on Multiparticle Dynamics}, edited by E. W. Kittel,
%		 W. Metzger, and A. Stergiou (World Scientific, Singapore, 1982), p. 964.
%
%\bibitem{fol01} L. L. Foldy and J. D. Walecka, \textit{Ann. Phys.}
%      \textbf{54} (1969) 447.
%
%\bibitem{bia01} A. Bianconi, S. Jeshonnek, N.N. Nikolaev and B.G. Zakharov,
%      \textit{Phys.  Lett.} \textbf{B343} (1995) 13.
%
\end{thebibliography}
\end{document}